\documentclass[%
 aip,
 amsmath,amssymb,
 reprint,%
]{revtex4-1}

\usepackage{graphicx}% Include figure files
\usepackage{dcolumn}% Align table columns on decimal point
\usepackage{bm}% bold math
\usepackage[utf8]{inputenc}
\usepackage[T1]{fontenc}
\usepackage{mathptmx}
\usepackage{etoolbox}

\makeatletter
\def\@email#1#2{%
 \endgroup
 \patchcmd{\titleblock@produce}
  {\frontmatter@RRAPformat}
  {\frontmatter@RRAPformat{\produce@RRAP{*#1\href{mailto:#2}{#2}}}\frontmatter@RRAPformat}
  {}{}
}%
\makeatother
\begin{document}
\preprint{AIP/123-QED}

\title[Sample title]{First observation of shock waves induced by laser-accelerated proton beams}
% Force line breaks with \\
\author{Yanlyu Fang}
\affiliation{State Key Laboratory of Nuclear Physics and Technology and CAPT, Peking University, Beijing 100871, China}
\author{Xiaoyun Le}
\affiliation{Department of Applied Physics, School of Science, Beijing University of Aeronautic and Astronautic, Beijing 100083, China}
\author{Yang Yan}
\author{Chentong Li}
\author{Mingfeng Huang}
\author{Yiting Yan}
\affiliation{State Key Laboratory of Nuclear Physics and Technology and CAPT, Peking University, Beijing 100871, China}%
\author{Xueqing Yan}
\author{Chen Lin}\email{lc0812@pku.edu.cn}
 \homepage{https://faculty.pku.edu.cn/linchen/zh_CN/}
\affiliation{State Key Laboratory of Nuclear Physics and Technology and CAPT, Peking University, Beijing 100871, China}
\affiliation{Beijing Laser Acceleration Innovation Center, Huairou, Beijing 101400, China}
\affiliation{Institute of Guangdong Laser Plasma Technology, Baiyun, Guangzhou 510540, China}

\date{\today}% It is always \today, today,
             %  but any date may be explicitly specified

\begin{abstract}
We demonstrate, for the first time, that laser-accelerated protons can induce shock waves in materials. The ultra-short pulse width of laser-driven protons enables them to deposit energy instantaneously, leading to an intense thermodynamic effect that heats and pressurizes materials violently, thereby generating shock waves. In contrast, laser-accelerated electrons do not possess this capability. Our simulations and experiments reveal that the flow intensity of the proton beam, which includes information on both the proton number and pulse width, directly correlates with shock waves. This finding not only provides a new method for characterizing the high flow intensity of laser-driven protons but also expands their applications in studying extreme states of matter. 
\end{abstract}

\maketitle

\section{Introduction}

Laser accelerators arose from the advent of chirped pulse amplification (CPA)\cite{CPA} in the 1980s, which enabled the realization of ultra-short and ultra-intense laser technology. This breakthrough allowed lasers to operate on the femtosecond time scale and achieve peak powers in the TW to PW range. Consequently, significant progress has been made in the field of laser ion acceleration. 
In this process, the laser initially transfers energy to electrons, which then create a high-gradient charge separation field of TV/m, thereby accelerating the ions within tens of micrometer-scale length\cite{TNSA}. 
In recent decades, numerous research studies\cite{1,2} have demonstrated that the interaction of superintense lasers with solid targets can generate proton beams with a broad energy spectrum, short pulse width (several picoseconds at the source)\cite{ps}, and high particle numbers (up to $10^{13}$)\cite{charge}. The highest energy recorded for laser-accelerated protons has now reached 150 MeV\cite{150mev}.

Due to the compactness, high quality and high intensity, the novel laser-driven particle accelerators have a broad range of applications in the fields of proton radiography\cite{radiography,radiography1}, proton radiotherapy\cite{me,me1}, materials irradiation\cite{ir1,ir2,ir3}, astrophysics\cite{as,as1}, etc. 
It has been found that by varying the distance from the proton source to the irradiated object, it is feasible to modulate the intensity and density of the proton beam, thus creating different irradiation conditions. 
For example, when the distance is within millimeter, an ultrashort-pulse beam of energetic protons will heat material to a warm dense plasma state at several eV\cite{wdm}. 
If the distance is lengthened to the centimeter order of magnitude, the material can be raised to temperatures of thousands of degrees, thus giving possibilities to study materials under extreme conditions\cite{ir2,ir3}. 
When the distance is further increased to the order of meters, protons can no longer cause significant heating of the irradiated material, but it can be used for radiography or tumor therapy when combined with precise beam modulation\cite{me1}. 

The short pulse widths of laser protons also make them easier to generate high thermal deposition gradients, resulting in ultrasonic signals.
The detection of proton-induced acoustics (protoacoustics) has been used to reconstruct range/dose distribution in proton therapy\cite{penghao1,penghao2,penghao3,protoacoustic1,protoacoustic2,protoacoustic3}, particularly at the Bragg Peak, thereby enhancing treatment accuracy by preventing under-dosing or over-dosing of surrounding healthy tissues\cite{protherapy}.
%In addition to those two aspects, researchers wondered whether the pulse width of proton beams could be characterized. 
However, previous studies have suggested that the inability to fulfill stress confinement in ultrasound detection limits the characterization of short pulse width beams\cite{protoacoustic1,protoacoustic-limit}.
Stress confinement requires that the proton pulse width ($t_\mathrm{p}$) be less than the thermal stress relaxation time ($t_\mathrm{S}$), where $t_{\mathrm{S}} = \mathrm{Bragg Peak_{FWHM}} /v_\mathrm{acoustic}$.
For example, in the research\cite{protoacoustic-limit}, a Bragg Peak full width at half maximum (FWHM) was 0.3 mm, and the velocity of sound in water was 1500 m/s, so $t_\mathrm{S}$ was 200 ns. 
It meant energy-selected and attenuated proton-induced acoustics could not characterize the ultra-short pulse width of laser-accelerated protons. Nevertheless, the above calculations are based on the assumption that $v_\mathrm{acoustic}$ is the intrinsic speed of sound of the material. If the energy deposition is strong enough, it will cause the generation of a shock wave and thus is expected to break the stress confinement.

The transition from ultrasonic signals to shock waves is a critical step. 
Acoustic waves propagating through a material cause an elastic deformation, which is reversible. At this point the material is subjected to little stress and the stress-strain response is linear (Hooke's law). 
When the energy deposition is sufficiently intense, the material undergoes rapid heating and expansion, leading to stresses that exceed the material's yield strength. This nonlinear response results in the formation of plastic deformation and shock waves, which propagate through the material with higher intensity and velocity than acoustic waves.

Laser-driven protons, with their ultra-short pulse width, can achieve a flow intensity of  $10^{10} \rm W/cm^{2}$ in a single shot.
This instantaneous energy deposition will cause a super-intense thermodynamic effect, making materials heat and pressurize violently\cite{shockdet}.
It suggests that laser-driven protons hold potential in the field of shock compression, which hasn't been fully investigated before. 
Unlike the traditional methods of generating shock compression\cite{dy1,dy2}, which include static high pressure generated by diamond anvil cells \cite{diamond} and dynamic high pressure driven by high-power pulse loading or strong lasers\cite{dy3},
laser-accelerated protons offer a unique approach. Their ultrashort pulse width and high particle numbers enable transient pulse loading, depositing energy almost entirely within a small area of the material. 
This efficient energy transfer facilitates shock wave generation. 
Additionally, the flow intensity of protons is proportional to the shock wave amplitude, providing a direct correlation between the proton beam characteristics and the resulting shock wave intensity.

In this paper, for the first time, we measured laser-accelerated proton beam-induced shock waves and provided the scaling law of shock wave evolution. In turn, the detection of the thermodynamic effect caused by instantaneous energy deposition can be used to characterize the flow intensity of protons. Since the flow intensity contains information on both the proton number and pulse width, the characterization of which is more universal and accessible.
Controlled experiments on a tandem accelerator have also verified that only ultrashort laser-accelerated protons can produce shock waves. 
This innovative means of generating shocks is crucial for understanding their extreme thermodynamic effects and serves as a powerful tool for studying extreme states of matter\cite{MRE1,MRE2}.

\section{Simulation}
The thermal effect induced by a proton beam in a material can be calculated by Eq.~\ref{eq1}. Here, $P(r, z, t)$ is the power density distribution function of the proton source and $T$ is the temperature of materials; $\rho$, $C_p$ and $\lambda$, are materials density, specific heat capacity, thermal conductivity, respectively.

\begin{equation}
\rho(T) C_p(T) \frac{\partial T}{\partial t}=\lambda(T) \nabla^2 T+P(r, z, t)\label{eq1}
\end{equation}

After obtaining $\Delta T$ at each moment from Eq.~\ref{eq1}, the stress in the material $\sigma$ can be calculated from Eq.~\ref{eq2}\cite{model,model2,model3}, including the effects of the material's coefficient of thermal expansion  $\alpha$ and its modulus of elasticity $G$.
\begin{equation}
\sigma=G \cdot \alpha(T) \cdot \Delta T\label{eq2}
\end{equation}

\begin{figure*}[htbp]
\includegraphics[width=0.85\textwidth]{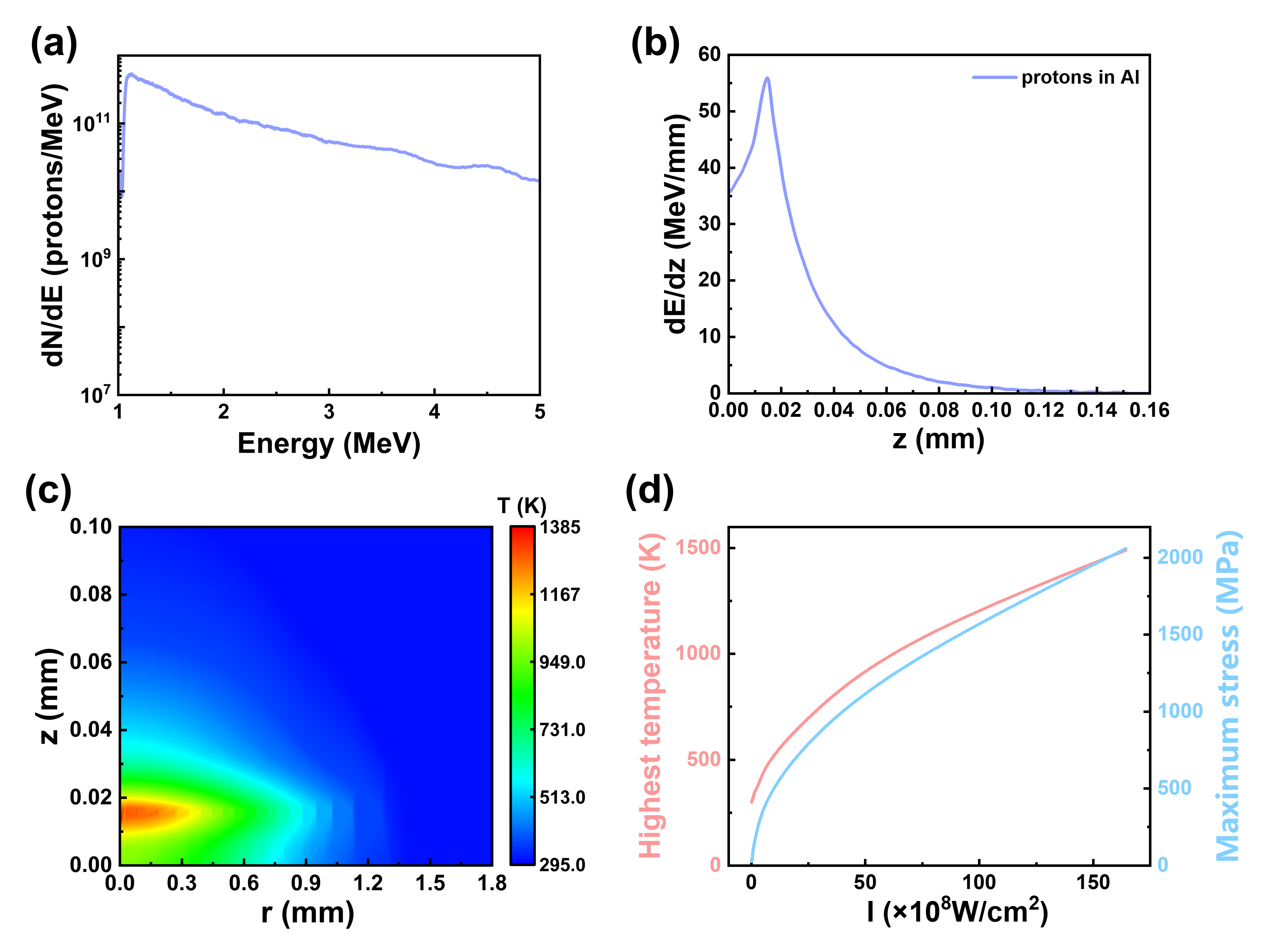}
\caption{Simulation results. (a) The proton energy spectrum. (b) The energy loss (dE/dz) curves in aluminum calculated by Geant4. (c) Temperature distribution in the radial and depth directions within the aluminum. (d) The variation of the highest temperature and stress in the material with proton flow intensity. }\label{fig3}
\end{figure*}

The key to the simulation is the modeling of the energy deposition process of laser-accelerated protons into a solid material, demoted as $P(r, z, t)$, considering their ultra-short pulse widths, broad energy spectra, and large scattering angles. 
%The energy spectral distribution and scattering angle affect the longitudinal and radial profiles of energy deposition, respectively. 
Fig.~\ref{fig3} (a) shows one energy spectrum of laser-driven proton beam acquired in the experiment, which satisfies an exponentially decreasing law\cite{pspectrum}:

\begin{equation}
\frac{\mathrm{d} N}{\mathrm{d} E}=\frac{N_0}{E} \exp \left(-\frac{E}{k_{\mathrm{B}} T}\right)\label{eq3}
\end{equation}  

The total energy deposited can be calculated from the energy spectrum:
\begin{equation}
E_{\text {total }}=\int_{E_{\min }}^{E_{\max }} E \cdot \frac{d N}{d E} \cdot d E\label{eq4}
\end{equation}

When the initial proton beam divergence angle is $\beta$, the beam spot radius after propagate distance $d$ can be calculated as:
\begin{equation}
 R_0=d \tan \beta \label{eq9}
\end{equation}

The pulse duration $\tau_b$ can be expressed as: 
\begin{equation}
\tau_b=d\left(\sqrt{\frac{m_p}{2 E_{\min }}}-\sqrt{\frac{m_p}{2 E_{\max }}}\right)\label{eq11}
\end{equation}
where $d$ is the distance between the target and sample, $m_p$ is the rest mass of the proton, and $E_{min}$ and $E_{max}$ are the minimum and maximum energy of the broadband proton beam. Therefor, in both simulation and experiments, we can control the pulse width, as well as the proton flow intensity, by controlling the irradiation distance $d$.

In the following article, the flow intensity of protons is:
\begin{equation}
I(d)=\frac{ E_{\text {total }}}{\pi R_0^2 \cdot \tau_b}\left[\frac{\mathrm{~W}}{\mathrm{~m}^2 }\right]\label{eq5}
\end{equation}
which includes the information of proton energy, radius $R_0$, and pulse width $\tau_b$.

Then its loaded power density in the material is:
\begin{equation}
P(r, z, t)=\frac{E_b}{\tau_b}=\frac{h(z) \cdot E_{\text {total }}}{\pi {R_0}^2 \cdot \tau_b}= h(z) \cdot I\left[\frac{\mathrm{~W}}{\mathrm{~m}^3}\right]
\label{eq6}
\end{equation}
where $E_b$ is the proton energy deposited in a unit volume, $h(z)$ is the normalized proton energy loss function of the broad-spectrum proton beam in the depth direction, $\tau_b$ is the pulse width, and $R_0$ is the proton beam spot radius.

Fig.~\ref{fig3} (b) uses Monte Carlo software Geant4\cite{geant4} to calculate the energy loss (dE/dz) curves of protons in Fig.~\ref{fig3} (a) in the material (aluminum). 
For a proton beam with a given energy distribution, the energy loss function is uniquely determined. Consequently, combining Eq.~\ref{eq1}, Eq.~\ref{eq2}, and Eq.~\ref{eq6}, the thermodynamic effect of the material is directly dependent on the proton flow intensity $I$.

Fig.~\ref{fig3} (c) shows the thermal effects induced in aluminum by such a broad-spectrum proton beam calculated with the finite element software Comsol Multiphysics. Currently, the irradiation distance is 0.5 cm and the flow intensity is more than $5\times10^{9} \ \rm{W/cm^{2}} $. 
It can be seen that the temperature distribution in the radial direction (r) is consistent with the proton radial distribution, while in the depth direction (z), it is also in agreement with the proton energy loss distribution in Fig.~\ref{fig3} (b). 
The temperature reaches its maximum at the Bragg peak position (less than 0.02 mm) inside the material, instantly forming a small but strong heat source. 
The material generates stresses due to thermal expansion, and the heat propagates as a wave, creating a shock wave. 
As the irradiation distance increases from 0.5 cm to 6 cm, the flow intensity decreases to $10^{8} \ \rm{W/cm^{2}}$ (see Fig.~\ref{SI-1} in the supplementary materials for the relationship between distance and flow intensity).
The variation of stress and temperature with flow intensity is shown in Fig.~\ref{fig3} (d).

It's worth noting that the rules of thermodynamic variations derived from the simulation are for reference only, and the specific parameters of the shock wave should still be based on the experiments. We aim to detect the shock wave experimentally and establish the correlation between the measured stress and the proton flow intensity.

\section{Experimental setup}

\begin{figure*}[htbp]
\centering
\includegraphics[width=0.9\textwidth]{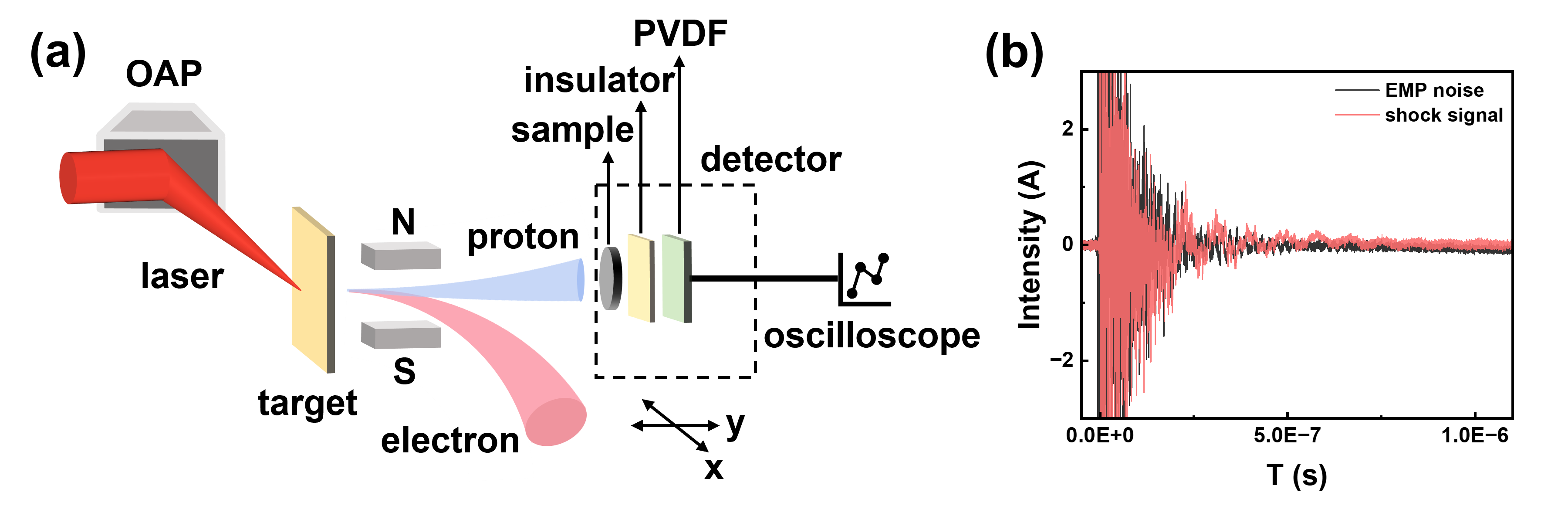}
\caption{(a) Experimental setup. (b) Original signals.}\label{fig1}
\end{figure*}

The experiments were performed on the compact laser-plasma accelerator (CLAPA) platform at Peking University\cite{CLAPA1, beamline1}. As shown in Fig.~\ref{fig1} (a), a p-polarized laser pulse with an energy of 1.3 J and a duration of 30 fs was focused by an f/3.5 off-axis parabola (OAP) to about 5 $\mu$m focal spot diameter (FWHM), corresponding to a laser intensity of $6\times10^{19} \ \rm{W/cm^{2}}$. The laser was incident onto a 7 $\mu$m thick aluminum foil at a $30^{\circ}$ incident angle to the target normal direction to accelerate protons. To diagnose protons, we employed a Thomson Parabola spectrometer, and the proton energy spectrum is shown in Fig.~\ref{fig3} (a).

The shock wave detector is placed behind the magnet, consisting of a sample, an insulating layer, a polyvinylidene difluoride (PVDF) piezoelectric film and a signal readout circuit. In order to detect proton-induced signals, a 0.7 T magnet was placed in front of the detector. 
The target sample was a millimeter-scale aluminum block. The instantaneous energy deposition of protons in the sample produced a strong thermodynamic effect, causing it to heat up dramatically, and at the same time generating a shock wave that transmitted into the material. When the wave propagated to the backside of the material, it gave a corresponding stress to the piezoelectric film causing an electrical charge on the surface and read out by an oscilloscope. The circuit section needs to be well-shielded. The entire detector and magnet were mounted on stepper motors. Transverse motor (x direction) movement controls whether the detector moves to its working position and receives the beam, while longitudinal motor (y direction) controls its detection distance.

When the detector is pushed away from the beam, it can only detect electromagnetic pulse (EMP) noise\cite{EMP}. When the beam is received, some low-frequency signals around 10 MHz clearly different from the EMP noise are measured, as shown in Fig.~\ref{fig1} (b), which are the shock wave signals generated by the laser protons. Based on the period of the wave, its velocity can be estimated to be 35 $km/s$, which is about six times greater than the intrinsic sound speed of aluminum.

\section{Results}
\begin{figure*}[htbp]
\centering
\includegraphics[width=0.85\textwidth]{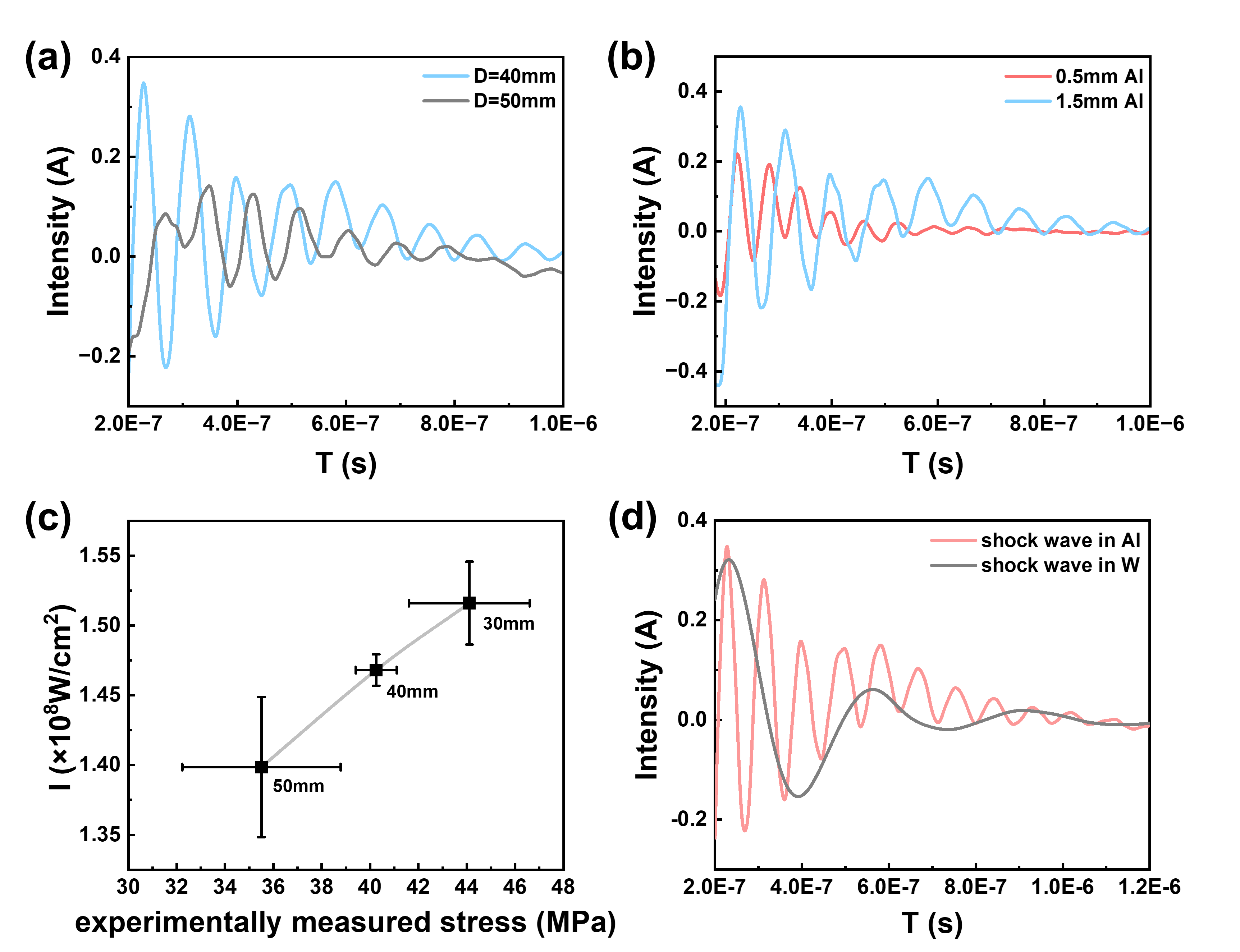}
\caption{(a) Comparison of shock wave signals at different detection distances. (b) Comparison of signals measured in samples of different thicknesses. (c) The relationship between stress and flow intensity of protons acquired from the simulation and experiments. (d) Shock waves detected in different materials.}\label{fig2}
\end{figure*}

To verify that the original signals in Fig.~\ref{fig1} were indeed shock wave signals, we evaluated them from two aspects:  

1. for the same sample, compare the signals at different detection distances $d$; 

2. for the same detection distance, compare the signals measured from different thickness samples.

In order to observe the changing rules of shock waves, we performed Fourier low-pass filtering on original signals. After removing the high-frequency noise, low-frequency signals were retained. As shown in Figure~\ref{fig2} (a), signals detected at different distances in the 1.5 mm thickness Aluminum sample are compared. With increasing irradiation distance, the proton flow intensity decreases, so the detected shock wave becomes weaker and irregular. 
On the other hand, at the same detection distance of 40 mm, the longer period of the shock wave is measured in the thicker sample (0.5 mm compared to 1.5 mm), as shown in Fig.~\ref{fig2} (b), which means the signal originates from a wave that reflects back and forth within the material. 
Combining these two points, it is verified that the signal is exactly the shock signal generated by laser protons in the material.

Taking the first period of d = 40 mm in Fig.~\ref{fig2} (a) as an example, after reducing the attenuation by a factor of 10 dB, the electricity signal is integrated to obtain the charge:
\begin{equation}
Q(t)=\int_0^t I(t) \mathrm{d} t \approx 1.2 \times10^{-8}(C)
\end{equation}
The stress on the PVDF can then be calculated from the charge.

\begin{equation}
\sigma(t)=\frac{1}{K_P A} Q(t)=\frac{1}{K_P A} \int_0^t I(t) \mathrm{d} t \approx 12(M P a)
\end{equation}
where $K_P$ is the piezoelectric coefficient and $A$ is the area of the piezoelectric film. 

Through this calculation method, we can obtain the stress in the material. 
In conjunction with the simulation, we can reconstruct the proton beam flow intensity incident on the sample.
Fig.~\ref{fig2} (c) shows the stress obtained from detecting the shock wave at different distances in the experiments along with the corresponding irradiation beam intensity.
It validates the feasibility of inferring the proton flow intensity by means of detecting shock waves.

We also compared the shock wave signals generated in different materials, such as aluminum and tungsten, under conditions of the same proton parameters and the same sample thicknesses, as shown in Fig.~\ref{fig2} (d). We can observe that the measured shock period in tungsten is longer than that in aluminum, indicating a slower wave velocity. In fact, the intrinsic sound speed in tungsten is also lower than that in aluminum. The measured shock wave velocity is consistent with existing research results\cite{Al-W}. One possible reason is that aluminum has more free electrons and better thermal conductivity, which enables it to transfer thermal energy to the surrounding medium rapidly, thereby making it easier to generate a strong shock.

\begin{figure}[htbp]
\centering
\includegraphics[width=0.46\textwidth]{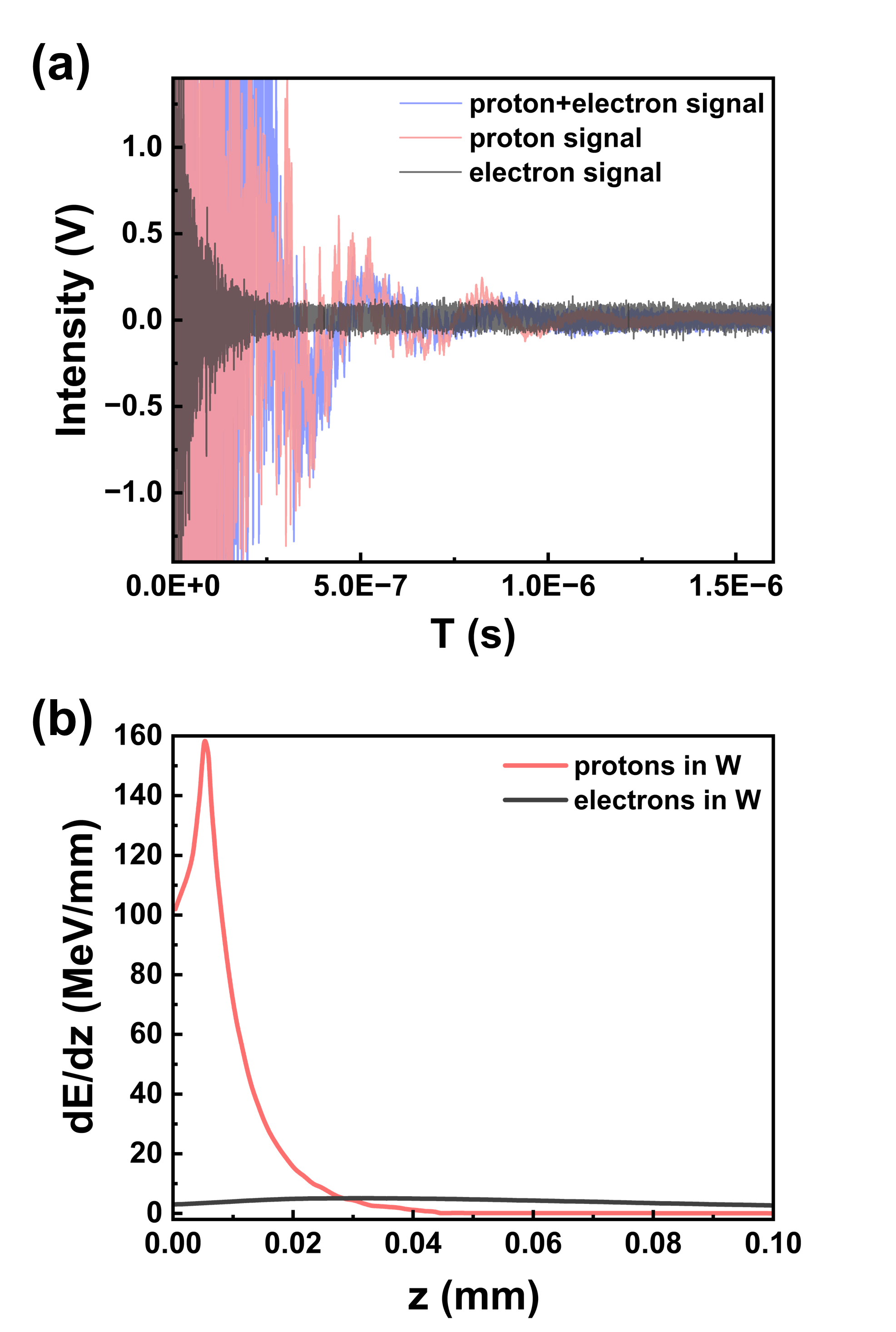}
\caption{(a) Shock waves generated by different particles in tungsten. (b) Energy loss distribution of protons and electrons with the same energy distribution in tungsten.}\label{fig4}
\end{figure}

In addition, we explored the shock signals generated by laser-accelerated electrons.
Aluminum cannot be used as a target material for detecting electron signals because electrons on the order of MeV can easily penetrate several millimeters of aluminum. We chose 1.5 mm tungsten, removed the magnet in front of the detector, and added a 200 $\mu m$ thick aluminum film to stop protons. 

The detection results are shown in Fig.~\ref{fig4} (a), where the energy deposition of electrons in tungsten does not generate a shock wave. There is only EMP noise in the measured signal, as shown by the black line. Considering it is possible that the high density and acoustic resistance of W make it unsuitable for detecting waves, we tested it with the experimental setup for protons and found that laser protons can produce a shock wave, as shown by the red line in Fig.~\ref{fig4} (a).
This is because electrons contribute to the overall heating of the material, unlike protons, which can create a small point heat source in the material. As shown in Fig.~\ref{fig4} (b), the red line and black line are energy loss distribution of protons and electrons with the same 1-5 MeV broad-energy spectrum distribution, calculated by Geant4. The energy loss value of protons is much higher than that of electrons, and has a Bragg peak effect.

\subsection{Shock wave detection on traditional accelerators}

\begin{figure}[htbp]
\centering
\includegraphics[width=0.5\textwidth]{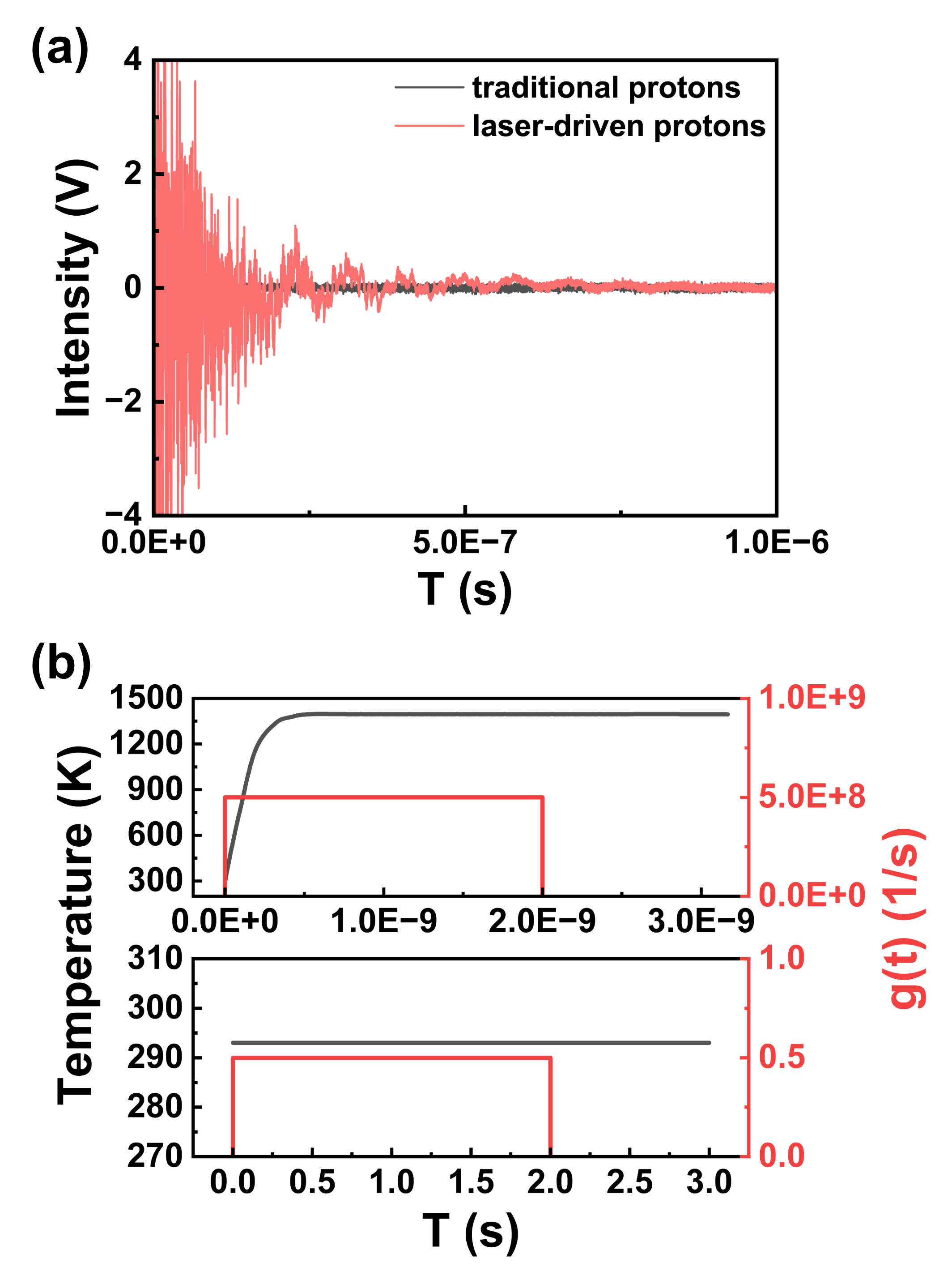}
\caption{(a) Simulated heating effect caused by proton beams with 2 ns and 2 s pulse widths. (b) Shock signals detected in experiments with two different proton beams.}\label{fig5}
\end{figure}

Furthermore, the shock wave detection experiment has also been conducted on the 1.7 MV tandem accelerator at Peking University\cite{1.7MV}. The detector did not trigger under uniform irradiation with a proton beam of 2 MeV energy, 200 nA current, and a beam spot size of 1 $cm^2$, as shown in Fig.~\ref{fig5} (a). The number of protons at that flow intensity was already far exceeding a single shot of laser-accelerated protons. Therefore it was verified that one of the necessary conditions for generating shock waves is a short pulse width. 

In order to explain and validate the above points, the heating effect caused by protons with different pulse widths was simulated, as shown in Fig.~\ref{fig5} (b). The upper graph corresponds to a 2 ns pulse width proton beam, which causes violent heating in the material (Al). The bottom graph is for 2 s pulse width and there is hardly any macroscopic temperature change.

This indicates only instantaneous thermal effect can produce a shock wave, which is the advantage of laser-driven particle beams.

Currently, laser-accelerated protons have reached a maximum energy of 150 MeV\cite{150mev}. The usage of protons with higher energies and charges would be expected to generate stresses up to the order of Mbar, thus enriching the realms of laser-driven protons in the field of shock waves, from astrophysics\cite{app}, geospace\cite{app1,app2}, medicine\cite{app4} to defense and military\cite{app3}.

\section{Conclusion}

In this paper, for the first time, it was demonstrated that laser-driven protons have the capability of generating shock waves owing to their ultra-strong transient energy deposition in matter, and that laser-driven electrons do not have this ability.
Therefore laser-driven protons will be a promising tool for rapidly generating shocks, affording us the remarkable opportunity to transcend the constraints of space and time to examine the behavior of various substances under extreme conditions.
The exploration of shock waves is a narrative of both scientific fascination and practical significance.

\section*{Acknowledgements}

We sincerely thank the 1.7 MV tandem accelerator staff, Professor Fu Engang and Xu Chuan for their precious support.
This work is supported by the National Grand Instrument Project (No. 2019YFF01014404), and the National Natural Science Foundation of China (Grants No. 12122501, No. 11975037, No. 61631001, and No. 11921006).

\section*{Author Declarations}
\subsection*{Conflict of interest}
The authors have no conflicts to disclose.

\section*{Supplementary materials}
\begin{figure}[htbp]
\centering
\includegraphics[width=0.38\textwidth]{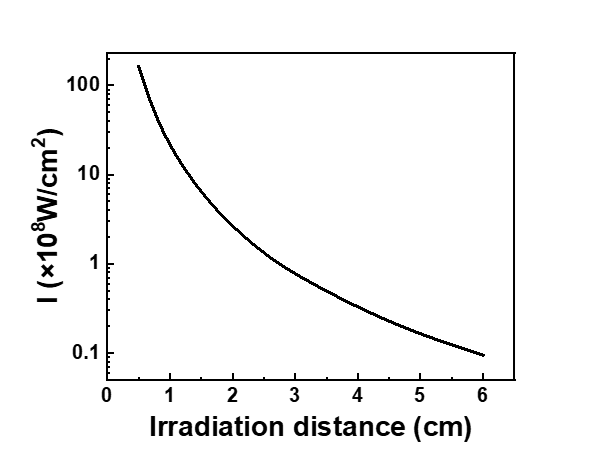}
\caption{The relationship between irradiation distance and flow intensity.}\label{SI-1}
\end{figure}

Fig.~\ref{SI-1} is the relationship between distance and flow intensity, which is mentioned in the simulation part.

\nocite{*}
\bibliography{aipsamp}% Produces the bibliography via BibTeX.

\end{document}